\documentclass[a4paper,reqno]{amsart}

\usepackage{amsfonts,amssymb}

\newtheorem{theorem}{Theorem}

\theoremstyle{remark}
\newtheorem{remark}{Remark}

\author{I.G. Korepanov}
\title[Algebraic relations for Pachner moves $3\to 3$ and $2\leftrightarrow 4$]{Algebraic relations with anticommuting variables for four-dimensional Pachner moves $3\to 3$ and $2\leftrightarrow 4$}
\date{}

\begin{document}

\begin{abstract}
Relatively simple algebraic relations are presented corresponding to Pachner moves $3\to 3$ and $2\leftrightarrow 4$, thus providing two thirds of the foundation for a four-dimensional topological quantum field theory. These relations are written in terms of Grassmann anticommuting variables and Berezin integral.
\end{abstract}

\maketitle

\section{Introduction}

The aim of this short paper is to present explicit solutions to four-dimensional analogues of pentagon equation, namely, algebraic relations corresponding to Pachner moves $3\to 3$ and $2\leftrightarrow 4$. Our relations are written in terms of Grassmann anticommuting variables and Berezin integral.

Recall that Pachner moves are elementary rebuildings of a manifold triangulation, and a theorem of Pachner states that any triangulation of a piecewise-linear (PL) manifold can be transformed into any other triangulation by a finite sequence of Pachner moves. Thus, an invariant of all Pachner moves is a PL manifold invariant and, moreover, one can hope to construct a topological quantum field theory on the base of algebraic relations corresponding to Pachner moves.

There are five kinds of Pachner moves in four dimensions: besides the mentioned $3\to 3$ and $2\leftrightarrow 4$, there are also moves $1\leftrightarrow 5$ which will be treated elsewhere.

Here we just present formulas that can be proved, e.g., using computer algebra. It will be the purpose of further and longer articles to explain the origin of these formulas. So, in a sense, we announce the existence of a theory leading to these formulas (and their generalizations). Right now we just note that these formulas come from a four-dimensional analogue of the scalar version of theory in~\cite{bk}, and that the four-dimensional analogue of the ``matrix'' version of that theory also exists and will be presented elsewhere.

\subsection*{Organization}

Below, we start in section~\ref{s:a} with reminding the basic definitions from the calculus of anticommuting variables. After this, we define in section~\ref{s:W} a weight factor corresponding to a four-dimensional simplex (4-simplex). In section~\ref{s:w} we define weights corresponding to inner two-dimensional faces (2-faces); this is done in a somewhat sophisticated way. Then, in section~\ref{s:33} we explain the Pachner move $3\to 3$ and present our algebraic relation corresponding to it, and in section~\ref{s:24} we do similar work for the move $2\to 4$. A formula for invariant of moves $3\to 3$ and $2\leftrightarrow 4$ is presented in section~\ref{s:i}. This formula and our tasks for further research are discussed in section~\ref{s:d}.

\section{Grassmann algebras, Berezin integral, and derivative w.r.t.\ anticommuting variables}
\label{s:a}

Recall that \emph{Grassmann algebra} over the field~$\mathbb C$ is an associative algebra with unity, having generators~$a_i$ and relations
\begin{equation}
a_i a_j = -a_j a_i, \quad \textrm{including} \quad a_i^2 =0.
\label{aa}
\end{equation}
Thus, any element of a Grassmann algebra is a polynomial of degree $\le 1$ in each generator~$a_i$, also called \emph{Grassmann variable}. The \emph{degree} of a monomial is its total degree in all Grassmann variables. If all the monomials in a polynomial are of degree~$n$, this latter is called \emph{homogeneous} polynomial of degree~$n$.

The \emph{Berezin integral}~\cite{B} is a $\mathbb C$-linear operator in a Grassmann algebra defined by equalities
\begin{equation}
\int da_i =0, \quad \int a_i\, da_i =1, \quad \int gh\, da_i = g \int h\, da_i,
\label{iB}
\end{equation}
if $g$ does not depend on~$a_i$ (that is, generator $a_i$ does not enter the expression for~$g$). Multiple integral is understood as iterated one, for instance,
$$
\int ab \,da\,db = \int \left( \int ab \,da \right) db = -1 .
$$

The (left) derivative $\partial / \partial a_i$ is also a $\mathbb C$-linear operator on Grassmann algebra. By definition, it acts on each monomial as follows: first, move the~$a_i$ to the left end of the monomial using commutation relations~\eqref{aa}, and then delete~$a_i$; if $a_i$ did not enter in the monomial, the derivative is zero.

\section{4-simplex weight}
\label{s:W}

We number triangulation vertices by natural numbers $1,2,\dots$, and also denote them by letters $i,j,\dots$ which are supposed to take values in natural numbers. Thus, every time we consider a simplex or a triangulation of something, the \emph{order} of vertices is fixed. For a simplex of any dimension, we write its vertices always in the increasing order, for instance, 2-face~$ijk$ has vertices $i$, $j$ and~$k$, with $i<j<k$.

To each vertex~$i$, we ascribe its \emph{coordinate} --- a complex number~$\zeta_i$. These numbers are arbitrary, with the only condition that $\zeta_i\ne\zeta_j$ if $i\ne j$. It will be convenient for us also to denote
$$
\zeta_{ij} \stackrel{\rm def}{=} \zeta_i - \zeta_j .
$$

Let there be some cluster of 4-simplexes --- by this we mean here that they form a triangulation of a manifold PL homeomorphic to the ball~$B^4$. To each tetrahedron~$ijkl$ --- a 3-face of one or two 4-simplexes --- let correspond two anticommuting variables $a_{ijkl}$ and~$b_{ijkl}$.

We are going to define the \emph{weight}~$W_{ijklm}$ of a 4-simplex~$ijklm$. We think, however, that the formula will be more readable if we write it with numbers rather than letters in subscripts. So,
\begin{multline}
W_{12345} \stackrel{\rm def}{=} \frac{1}{\zeta_{45}}
(\zeta_{34}a_{1234}-\zeta_{35}a_{1235}+\zeta_{45}a_{1245}-\zeta_{45}a_{1345}) \\
\cdot (\zeta_{34}b_{1234}-\zeta_{35}b_{1235}+\zeta_{45}b_{1245}+\zeta_{45}a_{2345}) \\
\cdot (-\zeta_{14}a_{1234}-\zeta_{24}b_{1234}+\zeta_{15}a_{1235}+\zeta_{25}b_{1235}-\zeta_{45}b_{1345}+\zeta_{45}b_{2345}) ,
\label{W}
\end{multline}
and $W_{ijklm}$ is obtained by the obvious substitution $1\mapsto i,\dots,5\mapsto m$.

The factor $1 / \zeta_{45}$ in fact cancels out in \emph{all} monomials obtained after expanding~\eqref{W}. So the weight~\eqref{W} is bilinear in~$\zeta$'s and, moreover, the coefficient at each product of $a$'s and/or $b$'s has the form $\zeta_{ij}\zeta_{kl}$, where the subscripts may coincide. The total number of such terms (monomials with nonzero coefficients~$\zeta_{ij}\zeta_{kl}$) in the expansion of~\eqref{W} is~72. The interested reader can see the expanded form of~$W_{12345}$ below in the Appendix.

\section{Weights for 2-faces}
\label{s:w}

We will also need weights for \emph{inner 2-faces} in a cluster of 4-simplexes. In contrast with weights~$W$ which will simply enter as factors in the integrand in our formulas \eqref{33} and~\eqref{24} below, the construction of a weight of several 2-faces involves a product of \emph{differential operators}, as we are going to explain.

First, we define the following differential operators for a two-face and a tetrahedron containing it. Like in formula~\eqref{W}, we prefer to put numbers rather than letters in subscripts, so we take tetrahedron~$1234$ and its four faces, having in mind that, for an arbitrary tetrahedron~$ijkl$ (remember that $i<j<k<l$), the substitution $1\mapsto i,\dots,4\mapsto l$ must be done. So, the operators are:
\begin{itemize}
	\item for face~$123$ and tetrahedron~$1234$,
\begin{equation}
\frac{\zeta_{23}}{\zeta_{34}}\frac{\partial}{\partial a_{1234}} - \frac{\zeta_{13}}{\zeta_{34}}\frac{\partial}{\partial b_{1234}}\, ,
\label{o123}
\end{equation}
	\item for face~$124$ and tetrahedron~$1234$,
\begin{equation}
-\frac{\zeta_{24}}{\zeta_{34}}\frac{\partial}{\partial a_{1234}} + \frac{\zeta_{14}}{\zeta_{34}}\frac{\partial}{\partial b_{1234}}\, ,
\label{o124}
\end{equation}
	\item for face~$134$ and tetrahedron~$1234$,
\begin{equation}
\frac{\partial}{\partial a_{1234}}\, ,
\label{o134}
\end{equation}
	\item for face~$234$ and tetrahedron~$1234$,
\begin{equation}
-\frac{\partial}{\partial b_{1234}}\, .
\label{o234}
\end{equation}
\end{itemize}

Second, let there be $n$ inner 2-faces in a cluster. For every such face~$ijk$, we define the differential operator~$d_{ijk}$ as the sum of all expressions of type \eqref{o123}--\eqref{o234} corresponding to \emph{all tetrahedra containing face}~$ijk$. Examples of such operators can be found below in formulas \eqref{dl33}, \eqref{dr33} and~\eqref{dr24}.

Third, we define multivalued operators~$d_{ijk}^{-1}$ as follows: let $f$ be a homogeneous element in the Grassmann algebra, then $g=d_{ijk}^{-1}f$ is any element satisfying $d_{ijk}g=f$ and also homogeneous, namely of degree $\deg g = \deg f + 1$. In the same way, we define the inverse to a product of $n$ differential operators:
$$
g= \left( \prod_{m=1}^n d_{i_mj_mk_m} \right)^{-1} f
$$
is any homogeneous element satisfying
$$
\left(\, \prod_{m=1}^n d_{i_mj_mk_m}\right) g = f \quad \text{and} \quad \deg g = \deg f + n .
$$

Finally, we construct the weight corresponding to the ordered $n$-tuple of 2-faces $i_1j_1k_1,\dots,i_nj_nk_n$ as any Grassmann algebra element to which the following expression can be equal:
\begin{equation}
w_{i_1j_1k_1,\dots,i_nj_nk_n} \stackrel{\rm def}{=} ( d_{i_1j_1k_1} \dots d_{i_nj_nk_n} )^{-1} 1 .
\label{prodd}
\end{equation}

\section{Move $3\to 3$ and the relation corresponding to it}
\label{s:33}

Consider the following $3\to 3$ move: the cluster of three 4-simplexes $12345$, $12346$ and~$12356$ is replaced by the cluster of simplexes $12456$, $13456$ and~$23456$. We also say that the initial cluster is in the ``left-hand side'' of the move (and the corresponding algebraic expression will stay in the l.h.s.\ of the relation~\eqref{33}), while the final cluster --- in the ``right-hand side''.

The boundary of both sides is of course the same: it consists of tetrahedra $1245$, $1246$, $1256$, $1345$, $1346$, $1356$, $2345$, $2346$, and~$2356$.

The \emph{inner} tetrahedra are, however, different: $1234$, $1235$ and~$1236$ in the l.h.s., and $1456$, $2456$ and~$3456$ in the r.h.s. Also, there is one inner 2-face~$123$ in the l.h.s., and one inner 2-face~$456$ in the r.h.s.

For the face~$123$, the corresponding differential operator is, according to \eqref{o123}--\eqref{o234} and the paragraph after these formulas,
\begin{equation}
d_{123} = \frac{\zeta_{23}}{\zeta_{34}}\frac{\partial}{\partial a_{1234}} + \frac{\zeta_{31}}{\zeta_{34}}\frac{\partial}{\partial b_{1234}} + \frac{\zeta_{23}}{\zeta_{35}}\frac{\partial}{\partial a_{1235}} + \frac{\zeta_{31}}{\zeta_{35}}\frac{\partial}{\partial b_{1235}} + \frac{\zeta_{23}}{\zeta_{36}}\frac{\partial}{\partial a_{1236}} + \frac{\zeta_{31}}{\zeta_{36}}\frac{\partial}{\partial b_{1236}} .
\label{dl33}
\end{equation}
So,
\begin{multline}
w_{123} = d_{123}^{-1} 1 \\
= \frac{p_{1234}a_{1234} + q_{1234}b_{1234} + p_{1235}a_{1235} + q_{1235}b_{1235} + p_{1236}a_{1236} + q_{1236}b_{1236}}{p_{1234}\frac{\zeta_{23}}{\zeta_{34}} + q_{1234}\frac{\zeta_{31}}{\zeta_{34}} + p_{1235}\frac{\zeta_{23}}{\zeta_{35}} + q_{1235}\frac{\zeta_{31}}{\zeta_{35}} + p_{1236}\frac{\zeta_{23}}{\zeta_{36}} +q_{1236}\frac{\zeta_{31}}{\zeta_{36}}}\, ,
\label{gwl}
\end{multline}
with any numbers $p_{\dots}$ and~$q_{\dots}$ such that the denominator in~\eqref{gwl} does not vanish. In particular, any one of the following can work as~$w_{123}$:
\begin{equation}
\frac{\zeta_{34}}{\zeta_{23}}\,a_{1234},\quad \frac{\zeta_{34}}{\zeta_{31}}\,b_{1234},\quad  \frac{\zeta_{35}}{\zeta_{23}}\,a_{1235},\quad \frac{\zeta_{35}}{\zeta_{31}}\,b_{1235},\quad \frac{\zeta_{36}}{\zeta_{23}}\,a_{1236}, \text{ \ or \ } \frac{\zeta_{36}}{\zeta_{31}}\,b_{1236} .
\label{6wl}
\end{equation}

Similarly, 
\begin{equation}
d_{456} = - \frac{\partial}{\partial b_{1456}} - \frac{\partial}{\partial b_{2456}} - \frac{\partial}{\partial b_{3456}} ,
\label{dr33}
\end{equation}
hence
\begin{equation}
w_{456} = \frac{p_{1456}a_{1456} + q_{1456}b_{1456} + p_{2456}a_{2456} + q_{2456}b_{2456} + p_{3456}a_{3456} + q_{3456}b_{3456}}{-q_{1456}-q_{2456}-q_{3456}}\, ,
\label{gwr}
\end{equation}
in particular, any one of the following can work as~$w_{456}$:
\begin{equation}
 -b_{1456},\quad   -b_{2456}, \text{ \ or \ } {-}b_{3456} .
\label{6wr}
\end{equation}

\begin{theorem}
\label{th:33}
The following identity, corresponding naturally to the $3\to 3$ move, holds:
\begin{multline}
\int W_{12345} W_{12346} W_{12356} w_{123} \frac{da_{1234}\,db_{1234}}{\zeta_{34}} \frac{da_{1235}\,db_{1235}}{\zeta_{35}}\frac{da_{1236}\,db_{1236}}{\zeta_{36}} \\
 = \int W_{12456} W_{13456} W_{23456} w_{456} \frac{da_{1456} \,db_{1456}}{\zeta_{56}} \frac{da_{2456} \,db_{2456}}{\zeta_{56}} \frac{da_{3456} \,db_{3456}}{\zeta_{56}} .
\label{33}
\end{multline}
\end{theorem}

Thus, the weights~$W$ in~\eqref{33} are taken for the 4-simplexes which are respectively in the l.h.s.\ and r.h.s.; the weights~$w$ are taken for the inner 2-faces; and the integration is performed in $da$ and~$db$ corresponding to the inner tetrahedra.

\begin{proof}
For this moment, it is enough to say that identity~\eqref{33} can be proved using computer algebra. We will give another proof, really showing \emph{why} \eqref{33} holds, in one of future papers.
\end{proof}

\begin{remark}
Our formulas do take into account the order of vertices. Nevertheless, if we number the vertices involved in the $3\to 3$ move in a different order, a formula of the same type as~\eqref{33} still holds. All these formulas are such that the expression~\eqref{ti} below stays invariant.
\end{remark}

\section{Move $2\to 4$ and the relation corresponding to it}
\label{s:24}

Here we consider the following $2\to 4$ move: the cluster of two 4-simplexes $12345$ and $12346$ is replaced by the cluster of four 4-simplexes $12356$, $12456$, $13456$ and~$23456$. The boundary of both sides consists of tetrahedra $1235$, $1236$, $1245$, $1246$, $1345$, $1346$, $2345$ and~$2346$.

In the l.h.s., there is one inner tetrahedron~$1234$ and no inner 2-faces.

In the r.h.s., there are six inner tetrahedra $1256$, $1356$, $1456$, $2356$, $2456$ and~$3456$, and four inner 2-faces $156$, $256$, $356$ and~$456$. The $d$-operators for these 2-faces are, according to \eqref{o123}--\eqref{o234}, as follows:
\begin{multline}
d_{156}=\frac{\partial}{\partial a_{1256}} + \frac{\partial}{\partial a_{1356}} + \frac{\partial}{\partial a_{1456}}\, , \quad
d_{256}= - \frac{\partial}{\partial b_{1256}} + \frac{\partial}{\partial a_{2356}} + \frac{\partial}{\partial a_{2456}}\, , \\
d_{356}= - \frac{\partial}{\partial b_{1356}} - \frac{\partial}{\partial b_{2356}} + \frac{\partial}{\partial a_{3456}}\, , \quad
d_{456}= - \frac{\partial}{\partial b_{1456}} - \frac{\partial}{\partial b_{2456}} - \frac{\partial}{\partial b_{3456}}\, .
\label{dr24}
\end{multline}
So, one can check that, for instance,
$$
a_{1256}b_{1256}a_{3456}b_{3456}
$$
is suitable as $w_{156,256,356,456}$ in formula~\eqref{24} below.

\begin{theorem}
\label{th:24}
The following identity, corresponding naturally to the $2\to 4$ move, holds:
\begin{multline}
\int W_{12345} W_{12346} \frac{da_{1234}\,db_{1234}}{\zeta_{34}} \\
 = - \zeta_{56} \int W_{12356} W_{12456} W_{13456} W_{23456} w_{156,256,356,456} \\
\cdot \frac{da_{1256} \,db_{1256}}{\zeta_{56}} \frac{da_{1356} \,db_{1356}}{\zeta_{56}} \frac{da_{1456} \,db_{1456}}{\zeta_{56}} \frac{da_{2356} \,db_{2356}}{\zeta_{56}} \frac{da_{2456} \,db_{2456}}{\zeta_{56}} \frac{da_{3456} \,db_{3456}}{\zeta_{56}} .
\label{24}
\end{multline}
\end{theorem}

Here, like in theorem~\ref{th:33}, the weights~$W$ in~\eqref{24} are taken for the 4-simplexes which are respectively in the l.h.s.\ and r.h.s.; the weight~$w$ is taken for the inner 2-faces; and the integration is performed in $da$ and~$db$ corresponding to the inner tetrahedra.

\begin{proof}
We refer again to computer calculations, with an intention to present in a future paper a proof showing real causes why \eqref{24} holds.
\end{proof}

\begin{remark}
The factor~$(-\zeta_{56})$ before the integral in the r.h.s.\ of~\eqref{24} is interpreted as corresponding to the \emph{inner edge}~$56$ in the cluster of four 4-simplexes. There are of course no inner edges in any cluster of two or three 4-simplexes considered in this paper.
\end{remark}

\begin{remark}
Again, like for move $3\to 3$, formulas of the same type as~\eqref{24} do hold for any other ordering of vertices.
\end{remark}

\section{Invariant of moves $3\to 3$ and $2\leftrightarrow 4$}
\label{s:i}

It can be checked using \eqref{33} and~\eqref{24} that the following expression, written for an arbitrary triangulated 4-manifold with boundary, remains invariant under moves $3\to 3$ and $2\leftrightarrow 4$:
\begin{equation}
\pm \prod_{\begin{smallmatrix}\text{over inner}\\ \text{edges }ij \end{smallmatrix} }
\, \int \!\!\!\!\! \prod_{\begin{smallmatrix}\text{over all}\\ \text{4-simplexes }ijklm \end{smallmatrix} } \!\!\! W_{ijklm}
\cdot w \cdot
\!\!\! \prod_{\begin{smallmatrix}\text{over inner}\\ \text{tetrahedra }ijkl \end{smallmatrix} } \!\!\! \frac{da_{ijkl}\, db_{ijkl}}{\zeta_{kl}}\, ,
\label{ti}
\end{equation}
where
\begin{equation}
w= \Biggl( \prod_{\begin{smallmatrix}\text{over inner}\\ \text{2-faces }ijk \end{smallmatrix} } d_{ijk} \Biggr)^{-1} 1 .
\label{tii}
\end{equation}
Here the sign~$\pm$ corresponds to the fact that we did not specify the exact order in the products; most likely, there exists some elegant formula relating this order and this sign.

\section{Discussion}
\label{s:d}

Formulas \eqref{ti} and~\eqref{tii} are of course related to Reidemeister-type torsions of some chain complexes, as it was, for instance, in our paper~\cite{bk}. They give a set of invariants of moves $3\to 3$ and $2\leftrightarrow 4$ in a compact and elegant form, uncovering, by the way, their ``fermionic'' nature, which manifests itself in the natural appearance of anticommuting variables.

The mentioned moves are, however, not enough for constructing invariants, and a topological quantum field theory, for four-dimensional manifolds with boundary: they do not change the number of inner vertices and, moreover, it can be shown that if there is at least one inner vertex in the triangulation, \eqref{ti} gives identical zero. This deternines the first item in the following list of tasks directly related to this paper:
\begin{itemize}
	\item extend our theory to all Pachner moves, including $1\leftrightarrow 5$,
  \item construct ``twisted'' versions of the theory in analogy with papers~\cite{M1,M2},
  \item generalize the theory to the matrix case in analogy with paper~\cite{bk},
  \item develop efficient calculation techniques and do calculations for various 4-manifolds with boundary and related objects.
\end{itemize}

% \appendix

\section*{Appendix}

Here is the expanded form of 4-simplex weight~$W_{12345}$, defined in compact form in formula~\eqref{W}:
\begin{multline*}
W_{12345} = \zeta_{34} \zeta_{45} a_{1234} a_{2345} b_{2345}-\zeta_{35} \zeta_{45} b_{1235} a_{1245} b_{1345}
+\zeta_{34} \zeta_{35} b_{1234} a_{1235} b_{2345}\\+\zeta_{15} \zeta_{45} a_{1235} a_{1245} a_{2345}
-\zeta_{15} \zeta_{35} a_{1235} b_{1235} a_{1345}-\zeta_{15} \zeta_{45} a_{1235} a_{1345} a_{2345}\\
-\zeta_{45}^2 a_{1245} b_{1245} b_{1345}-\zeta_{45}^2 a_{1345} b_{1345} a_{2345}
-\zeta_{13} \zeta_{35} a_{1234} a_{1235} b_{1235}\\-\zeta_{24} \zeta_{45} b_{1234} a_{1245} a_{2345}
-\zeta_{24} \zeta_{45} b_{1234} b_{1245} a_{1345}-\zeta_{13} \zeta_{34} a_{1234} b_{1234} a_{1235}\\
-\zeta_{34} \zeta_{23} a_{1234} b_{1234} b_{1235}-\zeta_{24} \zeta_{45} b_{1234} a_{1245} b_{1245}
+\zeta_{45} \zeta_{13} a_{1234} a_{1235} a_{2345}\\-\zeta_{14} \zeta_{34} a_{1234} b_{1234} a_{1345}
-\zeta_{14} \zeta_{45} a_{1234} a_{1245} b_{1245}+\zeta_{14} \zeta_{34} a_{1234} b_{1234} a_{1245}\\
+\zeta_{24} \zeta_{35} b_{1234} a_{1235} a_{2345}+\zeta_{45} \zeta_{13} a_{1234} a_{1235} b_{1245}
-\zeta_{25} \zeta_{45} b_{1235} a_{1345} a_{2345}\\-\zeta_{14} \zeta_{35} a_{1234} b_{1235} a_{1245}
-\zeta_{35} \zeta_{45} a_{1235} a_{2345} b_{2345}+\zeta_{35} \zeta_{45} b_{1235} a_{1345} b_{1345}\\
-\zeta_{34} \zeta_{15} b_{1234} a_{1235} a_{1345}+\zeta_{34} \zeta_{45} a_{1234} b_{1245} b_{2345}
-\zeta_{34} \zeta_{35} b_{1234} a_{1235} b_{1345}\\-\zeta_{45}^2 a_{1345} a_{2345} b_{2345}
-\zeta_{35}^2 a_{1235} b_{1235} b_{1345}+\zeta_{45}^2 a_{1245} a_{2345} b_{2345}\\
+\zeta_{35}^2 a_{1235} b_{1235} b_{2345}+\zeta_{34} \zeta_{24} a_{1234} b_{1234} b_{1245}
-\zeta_{34} \zeta_{45} a_{1234} b_{1245} b_{1345}\\+\zeta_{14} \zeta_{45} a_{1234} a_{1345} a_{2345}
+\zeta_{15} \zeta_{45} a_{1235} b_{1245} a_{1345}+\zeta_{45} \zeta_{23} b_{1234} b_{1235} a_{1345}\\
+\zeta_{35} \zeta_{45} b_{1235} a_{1245} b_{2345}-\zeta_{14} \zeta_{45} a_{1234} a_{1245} a_{2345}
-\zeta_{34} \zeta_{45} b_{1234} a_{1245} b_{2345}\\-\zeta_{34} \zeta_{25} a_{1234} b_{1235} a_{2345}
+\zeta_{25} \zeta_{45} b_{1235} a_{1245} a_{2345}-\zeta_{45}^2 b_{1245} a_{1345} b_{1345}\\
+\zeta_{45}^2 a_{1245} b_{1345} a_{2345}+\zeta_{45}^2 a_{1245} b_{1245} b_{2345}
-\zeta_{34}^2 a_{1234} b_{1234} b_{1345}\\+\zeta_{25} \zeta_{45} b_{1235} b_{1245} a_{1345}
-\zeta_{14} \zeta_{45} a_{1234} b_{1245} a_{1345}+\zeta_{24} \zeta_{35} b_{1234} a_{1235} b_{1245}\\
+\zeta_{35} \zeta_{25} a_{1235} b_{1235} a_{2345}-\zeta_{34} \zeta_{45} b_{1234} a_{1345} b_{1345}
-\zeta_{23} \zeta_{35} b_{1234} a_{1235} b_{1235}\\+\zeta_{34} \zeta_{45} b_{1234} a_{1245} b_{1345}
+\zeta_{35} \zeta_{45} a_{1235} b_{1245} b_{1345}-\zeta_{34} \zeta_{35} a_{1234} b_{1235} b_{2345}\\
+\zeta_{34} \zeta_{45} b_{1234} a_{1345} b_{2345}-\zeta_{35} \zeta_{45} a_{1235} b_{1245} b_{2345}
+\zeta_{15} \zeta_{35} a_{1235} b_{1235} a_{1245}\\+\zeta_{25} \zeta_{45} b_{1235} a_{1245} b_{1245}
+\zeta_{24} \zeta_{45} b_{1234} a_{1345} a_{2345}+\zeta_{15} \zeta_{45} a_{1235} a_{1245} b_{1245}\\
-\zeta_{45} \zeta_{23} b_{1234} b_{1235} a_{1245}+\zeta_{34} \zeta_{15} b_{1234} a_{1235} a_{1245}
+\zeta_{35} \zeta_{25} a_{1235} b_{1235} b_{1245}\\+\zeta_{34} \zeta_{24} a_{1234} b_{1234} a_{2345}
+\zeta_{34} \zeta_{45} a_{1234} b_{1345} a_{2345}+\zeta_{14} \zeta_{35} a_{1234} b_{1235} a_{1345}\\
-\zeta_{34} \zeta_{25} a_{1234} b_{1235} b_{1245}-\zeta_{35} \zeta_{45} a_{1235} b_{1345} a_{2345}
+\zeta_{34} \zeta_{35} a_{1234} b_{1235} b_{1345}\\+\zeta_{45}^2 b_{1245} a_{1345} b_{2345}
+\zeta_{34}^2 a_{1234} b_{1234} b_{2345}-\zeta_{35} \zeta_{45} b_{1235} a_{1345} b_{2345}\, .
\end{multline*}

\end{document}